# Efficient Integration of cross platform functions onto service-oriented architectures

Thomas Schulik[1], Viswanatha Reddy Batchu[1], Ramesh Kumar Dharmapuri[1], Saran Gundlapalli[1], Parthasarathy Nadarajan[1], Philipp Pelcz[1]

## 1 Abstract

The automotive industry is currently undergoing a major transformation with respect to the Electric/Electronic (E/E) and software architecture, driven by a significant increase in the complexity of the technological stack within a vehicle. This complexity acts as a driving force for Software-Defined Vehicles (SDVs) [1] leading to the evolution of the automotive E/E architectures from decentralized configuration comprising multiple Electronic Control Units (ECUs) towards a more integrated configuration comprising a smaller number of ECUs, domain controllers, gateways, and High-Performance Computers (HPCs) [2]. This transition along with several other reasons have resulted in heterogeneous software platforms such as AUTOSAR Classic [3], AUTOSAR Adaptive [4], and prototypical frameworks like ROS 2 [5]™[2]. It is therefore essential to develop applications that are both hardware- and platform/middleware-agnostic to attain development and integration efficiency. This work presents an application development and integration concept to facilitate developing applications as Software as a Product (SaaP), while simultaneously ensuring efficient integration onto multiple software architecture platforms. The concept involves designing applications in a hardware- and software platform-agnostic manner and standardizing application interfaces [6]. It also includes describing the relevant aspects of the application and corresponding middleware in a machine-readable format to aid the integration of developed applications. Additionally, tools are developed to facilitate semi-automation of the development and integration processes. An example application has been developed and integrated onto AUTOSAR Adaptive and ROS 2, demonstrating the applicability of the approach. Finally, metrics are presented to show the efficiency of the overall concept.

---

[1] All authors are from ZF Friedrichshafen AG. The authors have been listed alphabetically and have made equal contributions.

[2] ROS™, ROS 2™, and the “Nine Dots” logo are trademarks of Open-Source Robotics Foundation, Inc. (Open Robotics). The use of these marks is for descriptive purposes only and does not imply endorsement. All references to ROS and ROS 2 are made in accordance with the ROS Trademark Rules and Guidelines.

## 2 Introduction

Due to the increasing number of functions and the associated increase in complexity in distributed vehicle architectures, solutions are required which on one hand minimize the number of control units and on the other hand facilitate the efficient integration of functions in vehicles. This requires fundamentally new centralized vehicle architectures and technical solutions for flexible communication between the functions. This trend also requires that functions need to be decoupled from the underlying EE-Architecture, and they shall be implemented as independent services. This enhances the flexibility and efficiency in deployment. Furthermore, the effort to integrate functions as services and costs are reduced. This leads with well-known DevOps cycles to a faster development and maximum reuse of existing functions.

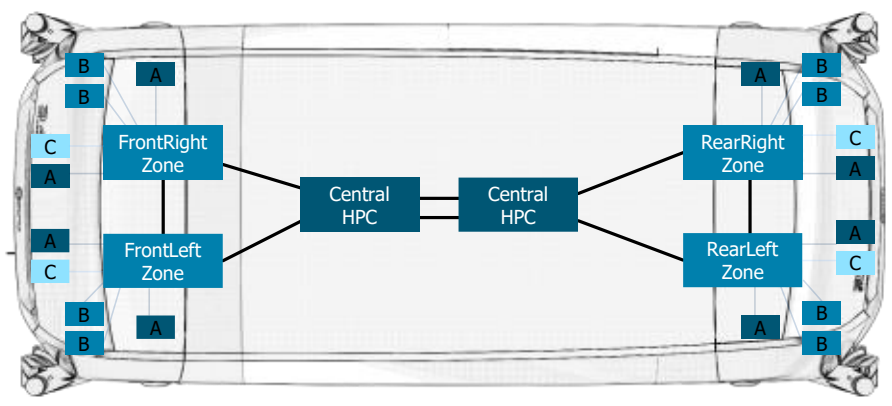

*Figure 1: Schematic view of a vehicle with centralized architecture divided into four zones (Front Left, Front Right, Rear Left, Rear Right), each containing three components labelled A, B, and C, all connected via lines to two central high-performance computers.*

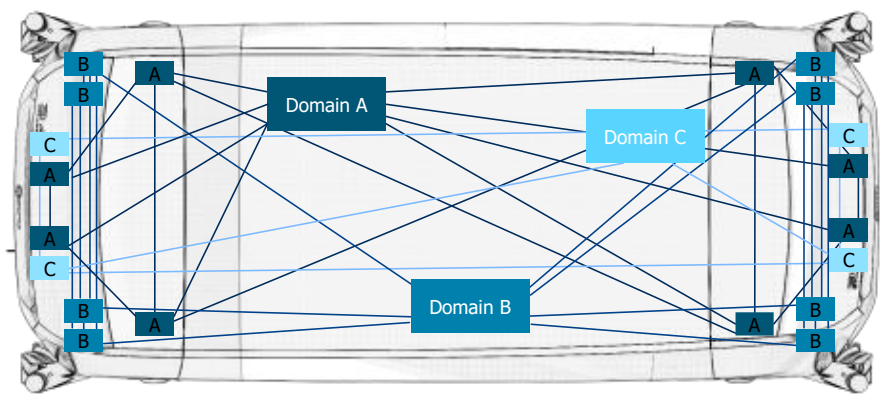

*Figure 2: Distributed vehicle architecture: Top-down schematic of a vehicle showing three internal domains (Domain A, Domain B, Domain C), each connected via blue lines to multiple labelled points (A, B, C) distributed around the vehicle's perimeter.*

## 3 Problem Statement

To support the exchange of functions between different suppliers and platforms, it is essential to decouple functions from the underlying ecosystem. The hardware component-wise integration of features in vehicles is cumbersome, as the EE-Architecture, in the past mostly based on CAN bus technology, needs to be adapted. Additionally, the reusability of these hardware components is limited due to system-specific, for example CAN signal to function interface mapping. Signal definitions (naming, resolution, frequency) also vary among vehicle manufacturers, impacting safety-related functions and architecture.

Functions based on AUTOSAR Classic have been tightly coupled to the APIs of this framework. Development tools, while efficient for the selected framework and vendor, lack compatibility to address the challenges of the emerging trend of Software-defined Vehicles (SdV). For security and proprietary reasons, vehicle signals in passenger vehicles exchanged on the CAN bus are not standardized.

We propose a concept to decouple functions through the harmonization of APIs and signals, leading to new requirements for software architecture, processes, and tools. Model-based tools can enhance the development process of modern software functions but require a definition of exchangeable formats to improve handover in software development. We suggest using a common model, format definition, and tools to enable the reuse of information at the software level design.

While the underlying communication technology (Automotive Ethernet) is an enabler for service-based architecture, its configuration is not within the scope of this work. Similarly, middleware development is excluded. Our primary goal is to outline the method, process, and required tools to demonstrate a reusable and efficient integration of functions as services. This approach aims to facilitate reuse in consortia like Eclipse-SdV [7] and other funded projects. The development of open-source tools is a crucial aspect, and we contribute with our process, methods, and model definitions.

## 4 Methodology

The following chapters aim to elaborate on the various requirements outlined in the introduction and problem statement. In Chapter 4.3, we address the challenge of creating a platform-independent function by introducing a defined structure for function calls. Chapter 4.6 enhances the concept from the COVESA VSS [8] project to standardize communication signals. This chapter also describes the development process and illustrates the tools involved. Chapters 4.10 and 4.11 present models for describing interfaces and platform requirements. In Chapter 5, the artifacts from the previous chapters are integrated to generate code in a semi-automated process, enabling the function to be deployed across various platforms. It is important to note that the validation of a function on a combination of hardware, operating system, and middleware is not considered here; the focus is solely on the requirements and integration. An example application is integrated into multiple runtime environments, such as AUTOSAR Adaptive and ROS 2, to address as many requirements as possible, serving as a reference implementation. Finally, runtime measurements are conducted and evaluated to assess the ability to integrate a function into different platforms.

Figure 3 shows a structured approach using model templates represented using the Pydantic [9] library to define both the function model and integration model, enabling strict data validation and standardization. Function requirements together with standardized interfaces are first used to generate a Function JSON, which contains function-related details while remaining independent of any middleware-specific dependencies. Middleware information is then applied to enhance the Function JSON, transforming it into Integration JSON, which incorporates necessary middleware-specific configurations. Finally, a Function Adapter, a dedicated code module is used to connect the function with the underlying middleware, ensuring efficient execution and integration. This approach promotes automation, reusability, and consistency in function development.

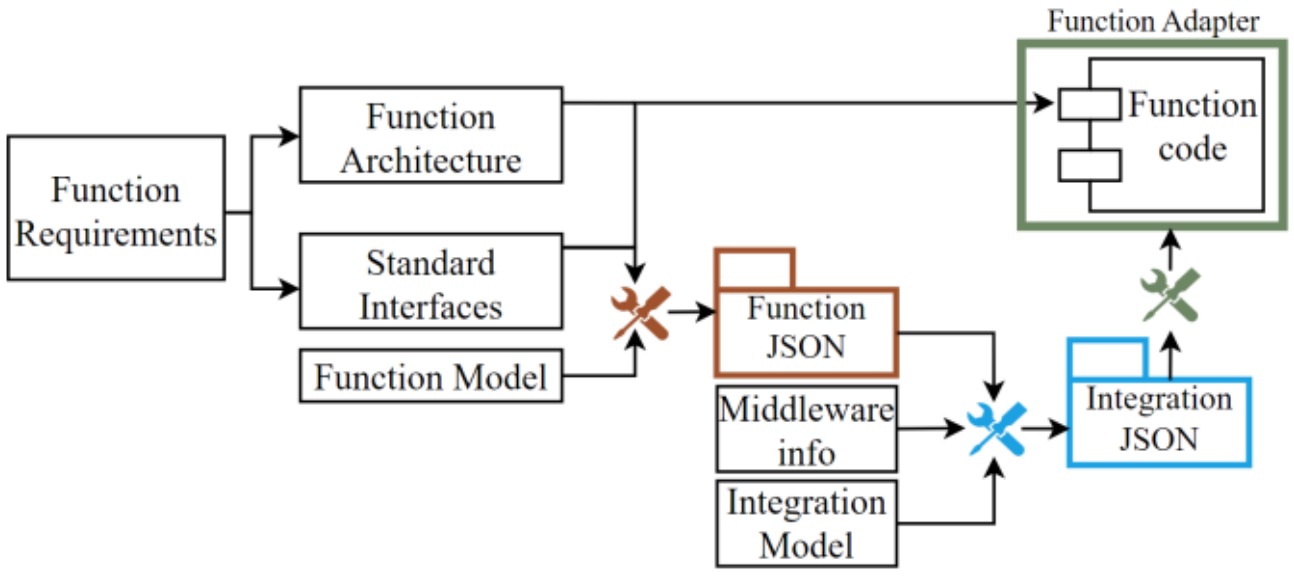

*Figure 3: Proposed function integration workflow.*

A key concern in most of the application functions is that the requirements that correspond to core functionality and project or platform specific needs doesn't evolve isolated but often get overlapped, devoid of modularity. And for several such applications, functional part that depends on platform/base software (BSW) features, doesn't hold proper isolation between layers.

As shown in Figure 4, overlap of different functional aspects from various requirements (core, platform/project specific etc.) is quite evident and the essential solution is ensuring that those application functions, which may vary in requirements across platforms, can be split, standardized, and integrated without compromising performance or compatibility.

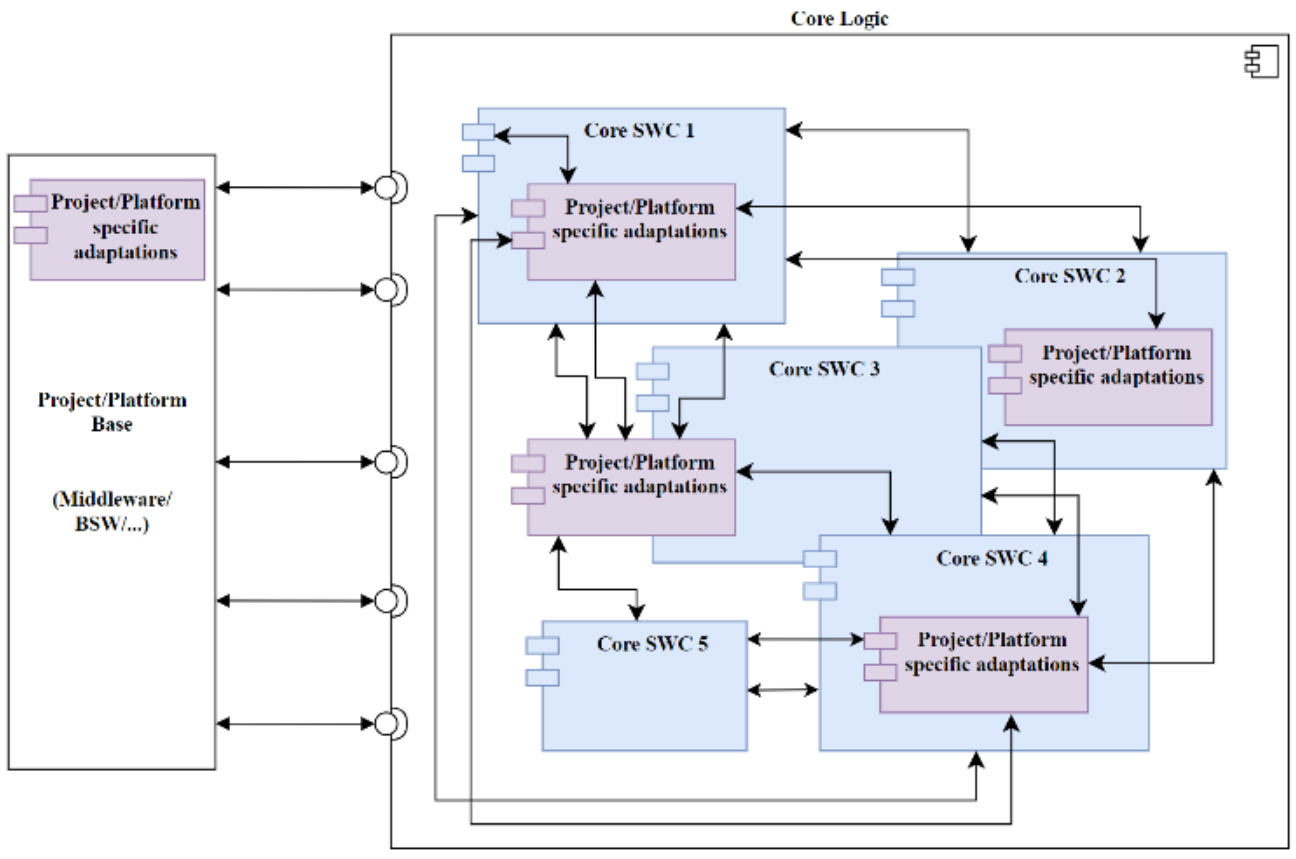

*Figure 4: Current view of a monolithic function architecture.*

This chapter addresses these challenges by discussing approaches to modular design of application functions. Specifically, it focuses on the advantages of dividing application functions into core modules and project - or platform-specific modules, with standardized APIs enabling seamless integration and customization across platforms.

Using a layered architecture as shown in Figure 5, where the interfaces form a boundary between core and platform-specific modules and a boundary between application and base software, will enable easy addition of new features without modifying existing core functionality. For instance, developers can add pre-processing or post-processing blocks to manage additional data transformations; multiplexing, demultiplexing, remapping or transforming error handling information [10]; or routing the logging data as required by a particular platform. This flexible architecture allows platform-specific customizations to be implemented as separate modules that interact with the core through well-defined APIs.

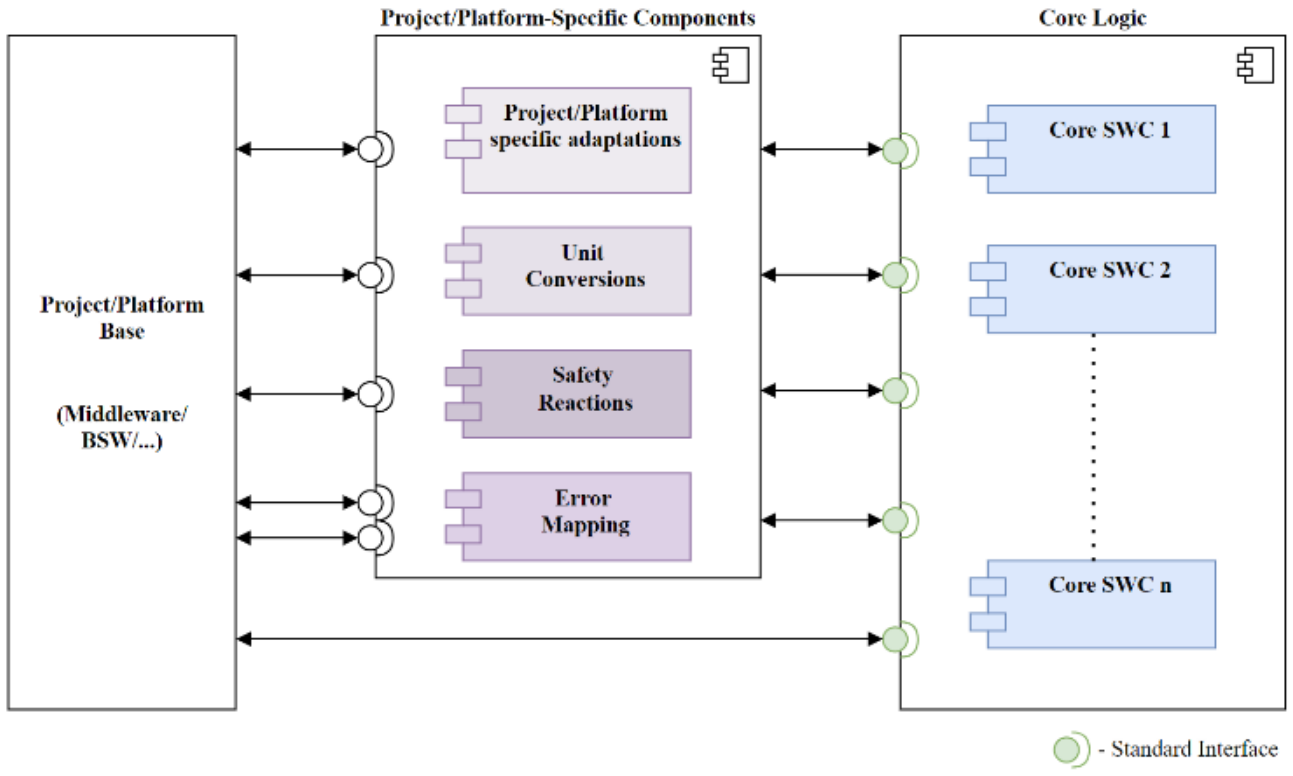

*Figure 5: Proposed modularized platform function architecture.*

As part of the component design, each function is structured with `init`, `step`, and `terminate` methods to handle different stages of its execution flow in a systematic manner. The `init` method is responsible for tasks such as signal initialization and memory reads when the component is first activated. The `step` method executes the core functionality during runtime, including periodic computations and data processing. Finally, the `terminate` method ensures proper shutdown behavior by handling activities such as memory write requests and resource cleanup. This separation of concerns improves reliability, maintainability, and modularity in the overall execution flow of the component

## 4.1 Core and Project/Platform-Specific Modules

To manage cross-platform functionality efficiently, the application function should be divided into core and project-specific or platform-specific modules. This approach is based on modularity principles, which advocate separating general-purpose logic (the "core" module) from platform-specific implementations.

## 4.2 Core Module Design:

The core module should contain all generic functionality that does not depend on platform-specific features. It includes the main algorithms, business logic, and other reusable components that can be used across multiple platforms. By abstracting platform-specific details from the core, developers create a codebase that is resilient to changes in underlying hardware or operating systems. This core module shall be designed to be stable, robust, and independent of any unique attributes of the deployment environment.

## 4.3 Project/Platform-Specific Modules:

These modules handle all unique aspects of a particular project or platform, such as specific I/O operations, hardware interfaces, or performance optimizations tailored to specific ECU or HPC configurations. They act as an extension to the core module, adding additional functionality or pre/post-processing blocks needed for each platform. This separation allows platform-specific requirements to be integrated without altering the core functionality, preserving the integrity and reusability of the main application.

## 4.4 Standardization of interfaces

To bridge the core and platform-specific modules, and to make application functions easily portable on to different platforms/BSWs, standardized APIs serve as connectors that enable communication and data exchange between these layers. The APIs provide a unified interface for the core module to interact with the platform-specific components, and a unified interface for core/platform specific components interact with BSW block set, ensuring a consistent communication structure.

API design should prioritize consistency and compatibility across platforms. By standardizing APIs, developers can ensure that platform-specific modules can plug into the core module with minimal modifications and the application function i.e., combination of core and platform-specific components can be integrated onto any platform with trivial changes and fewer integration steps. These APIs should encapsulate platform-specific details, presenting a unified interface to the core module while allowing the underlying implementation to vary based on platform requirements Figure 5.

COVESA VSS provides a structured framework for defining vehicle signals and data points (information origins such as sensor readings, control signals etc.) in a standardized way, facilitating the interaction between software modules across different ECUs and HPC platforms. VSS defines a hierarchical and extensible data model that allows OEMs (Original Equipment Manufacturers), suppliers, and software developers to agree upon a unified set of vehicle signals. VSS simplifies cross-platform integration by defining a universal "language" for vehicle data. By adhering to VSS, software functions can leverage consistent, well-defined signal names and structures, reducing ambiguity and enhancing compatibility between ECUs and HPCs.

### 4.5 Standardization and API Alignment with VSS:

VSS standardizes not only signal definitions but also data types, data groups (arrays, strings etc.), units, and hierarchical organization of vehicle data. When implementing standardized APIs, developers can align API calls and data structures with the VSS data model, ensuring that data exchanged between the core and platform-specific modules adheres to this shared framework.

### 4.6 Extension of standardization for non VSS interfaces:

Since VSS focuses solely on data and parameter interfaces, additional interfaces for error handling, safety reactions, mode management, scheduling, and watchdog functions are standardized separately. These interfaces are defined based on the configuration requirements of various Basic Software (BSW) modules, ensuring that each interface includes the essential fields needed for seamless integration with any BSW.

### 4.7 Standard interface definition:

**Data/Parameter Interfaces:** In VSS, branching is done through a hierarchical tree structure that organizes vehicle data signals. Each branch represents a different category or subsystem of the vehicle, such as powertrain, chassis, or infotainment. Within these branches, signals are further divided into subcategories, allowing for a clear, organized structure where each signal is uniquely identified based on its position in the hierarchy. This branching approach enables intuitive navigation and scalability, allowing new signals or branches to be added easily without disrupting existing structures.

This hierarchical branching is directly reflected in the naming convention, which uses dot-separated paths to indicate each level of the hierarchy. For example, a signal representing speed of the front-left wheel might be named `Vehicle.Chassis.Axle.Row1.Wheel.Left.Speed`. Each part of the name (`Vehicle`, `Chassis`, `Axle`, `Row1`, etc.) represents a branch in the hierarchical structure, making the location and context of each signal clear. This naming approach helps standardize the signal paths across different platforms, simplifying data integration, readability, and reusability.

Additional interface categories are standardized to manage the functions beyond data and parameter handling. These interfaces cover areas such as error management, safety reactions, mode management, scheduling, and watchdog functionality, each defined to meet specific system requirements and ensure compatibility with the BSW layer.

### 4.8 Core Module Using Standardized Interfaces:

By adopting standardization, the core module can be developed in a way that is entirely platform-agnostic. Since the mentioned formats standardizes the signal names, structure, and data format, the core module can process them without needing to know the specifics of the underlying hardware or ECU architecture. This abstraction makes the core more reusable and compatible across a variety of platforms that also adhere to this approach.

### 4.9 Platform-Specific Modules Extending VSS Standardization:

In cases where platform-specific adaptations are necessary, respective modules can introduce additional processing that interprets or transforms the signals in ways tailored to the requirements of a specific ECU or HPC. For instance, a platform-specific module might be implemented as a pre- or post-processing block that adapts VSS signals to specialized sensors, network protocols, or data logging systems unique to a specific vehicle model or region. Furthermore, these modules can modify, or route information related to error, mode, or safety reaction signals that are sent to or received from the BSW. Additionally, during function integration, all information encapsulated within each interface should be considered to configure or adapt the relevant middleware or BSW without requiring any modifications to the core application function modules.

### 4.10 Function model

Since the functions should be independent of hardware, underlying base software or middleware, and source code language (e.g., C/C++), we need a standardized model. This model will serve as a reference for projects, helping to define and understand the function interfaces such as interface data, parameters, errors, safety reactions, etc. along with their related configurations derived from software requirements. The function model shall be the primary source for referencing standard platform functions. It shall abstract the application in a way that allows integration with the underlying base software. This model can be designed in various formats, such as XML or JSON. Below is an example of how a function model is represented using the Pydantic [9] library, which can also generate the corresponding JSON schema.

```
class Function(BaseModel):
    Name: str
    Description: str
    InterfaceData: List[InterfaceDatatype]
    SchedulingInfo: SchedulingInfo
    MessageList: Optional[List[Message]]
    MethodList: Optional[List[Method]]
    ParameterList: Optional[List[Parameter]]
    ErrorList: Optional[List[Error]]
    SafetyReactionList: Optional[List[SafetyReaction]]
    Watchdog: Optional[Watchdog]
    AllocationInfo: Optional[AllocationInfo]
```

*Figure 6: Overview of the Function model*

```
class InterfaceDatatype(BaseModel):
    Name: str
    Description: str
    Role: Annotated[str, StringConstraints(pattern=r'^(Consumer|Provider)$')]
    Type: str
    Datatype: Union[Numerical, StringDatatype, BooleanDatatype, StructDatatype]
    RangeErrorAction: Optional[Annotated[str, StringConstraints(pattern=r'^(Default|Init)$')]]
    TimeoutValue: Optional[float]
    TimeoutError: Optional[str]
    AsilInfo: AsilLevel
```

*Figure 7: Overview of the Interface Datatype in Function model*

```
class Parameter(BaseModel):
    Name: str
    Description: str
    AsilInfo: AsilLevel
    Datatype: Union[Numerical, StringDatatype, BooleanDatatype, StructDatatype]
    RangeErrorAction: Optional[Annotated[str, StringConstraints(pattern=r'^(Default|Init)$')]]
    Attribute: Annotated[str, StringConstraints(pattern=r'^(NA|Normal|LearningParameter)$')]
```

*Figure 8: Overview of the Parameters in Function model*

For more details on individual sections of the function model such as SchedulingInfo, Watchdog, and others please refer to the complete Pydantic Function Model (fdk_function_model.py), the corresponding JSON Schema (function_json_schema.json), and the Function JSON for the CoreAcc platform function (CoreAcc_Function.json) available in the artifacts[3].

[3] https://github.com/ZF-Group/ELIV-presentation/tree/main

## 4.11 Integration model

As the *Function model* is agnostic to the underlying software platform, it is necessary to provide additional information whenever the function is to be integrated on a specific software platform. This information is required to create and manage software components and their corresponding configurations with respect to the software platform. For example, consider the AUTOSAR Adaptive software platform. When a function needs to be integrated on this platform, it is necessary to specify the adaptive software components along with their services and their corresponding interfaces. The services between the software components must be mapped and the service instances must be configured accordingly. It is also necessary to define and configure the executables. In addition, communication mechanisms such as SOME/IP, diagnostics, persistency, safety, and security mechanisms must also be configured in accordance with the relevant specifications.

With multiple software platforms emerging in the market, it is advisable to describe the specifications or requirements in a standardized manner for each software platform. This approach has been adopted in this work to develop the *Integration Model*. The Integration Model is a standardized model for describing the integration specification of an application with respect to an underlying software platform. It is crucial to emphasize that the Integration Model is inherently linked to the software platform, necessitating the generation of a unique model for each platform. The specifications, as with the Function Model, are provided in the JavaScript Object Notation (JSON) format [11]. Therefore, it is necessary to create JSON schemas to ensure standardization of the models.

The Function Model along with software platform act as input to the generation of the Integration Model. An exemplary Integration Model for Adaptive AUTOSAR is seen in the Figure 9.

```
class Component(BaseModel):
    Name: str
    ExecutableName: str
    FunctionGroupModes: List[str]
    FunctionList: List[Function]
    ParameterList: Optional[List[Parameter]]
    ServiceInterfaceList: List[ServiceInterface]
class ApplicationModel(BaseModel):
    MetaInformation: MetaInformation
    ApplicationInformation: ApplicationInformation
    ComponentList: List[Component]
    DataTypes: Optional[List[Union[TypeReference,
                                   StructDatatype,
                                   ArrayDatatype,
                                   EnumerationDatatype]]]
```

*Figure 9: Example Integration Model for Adaptive AUTOSAR.*

For the sake of simplicity, only the `ApplicationModel` that contains the list of software components, data types, and the `Component` that contains the list of functions, parameters, and service interfaces are shown. The generated Integration JSON file then serves as input to the function adapter, which is detailed in the following section.

### 4.12 Concept of function adapter

To systematically integrate functions into a middleware framework, the idea is to utilize a unified integration approach for each non-functional requirement. This means that each requirement is implemented step-by-step through the available APIs of a specific middleware framework. The entirety of the code for middleware integration of a function is then called a function adapter. The platform function is embedded as a class attribute in the function adapter. Thus, the function adapter integrates a function into a specific middleware while considering the non-functional requirements. Consequently, there is a separate function adapter for each middleware. A unified and middleware-framework-independent implementation of the function adapter is currently not feasible due to the different functionalities and APIs of the middleware frameworks.

## 5 Case Study

### 5.1 Setup description

To demonstrate the applicability of the approach, the workflow has been applied to an Advanced Driver Assistance Application (ADAS) application named as EcoControl for Powertrain. The application consists of two components namely Core Adaptive Cruise Control (Core ACC) and EcoControl Model Predictive Control (EcoControl MPC). The artifacts for the EcoControl application are generated with respect to two different software platforms namely Adaptive AUTOSAR and ROS 2. The overall architecture of the case study is illustrated in Figure 10.

An open loop setup has been established to validate the approach and derive the necessary metrics. CAN messages required for the Core ACC component are replayed via CANalyzer. These CAN messages are then converted to Scalable service-Oriented Middleware over IP (SOME/IP) or Data Distribution Service (DDS) messages, depending on the corresponding software platform used (SOME/IP for Adaptive AUTOSAR and DDS for ROS 2) in the Gateway hardware (NXP S32G [12]). An ethernet switch then transmits the converted messages to the Core ACC and EcoControl MPC. The Core ACC and EcoControl MPC run on the HPC (DATALynx ATX2 [13]). The Operating System (OS) on the Gateway is an Yocto Project®[4] of Linux®[5], while the HPCs run Ubuntu®[6] 22.04. It should also be noted that the applications run inside dedicated Docker containers [14].

---

[4] Yocto Project® is a registered trademark of the Linux Foundation. The use of the name is for descriptive purposes only and does not imply endorsement by the Linux Foundation or the Yocto Project. All references to the Yocto Project are made in accordance with its trademark usage guidelines.

[5] Linux® is the registered trademark of Linus Torvalds in the U.S. and other countries. The use of the term "Linux" in this publication is for descriptive purposes only and does not imply endorsement by Linus Torvalds or the Linux Foundation.

[6] Ubuntu® is a registered trademark of Canonical Ltd. The use of the name "Ubuntu" in this publication is for descriptive purposes only and does not imply endorsement by or affiliation with Canonical Ltd.
All references to Ubuntu are made in accordance with Canonicals' trademark and brand usage guidelines.

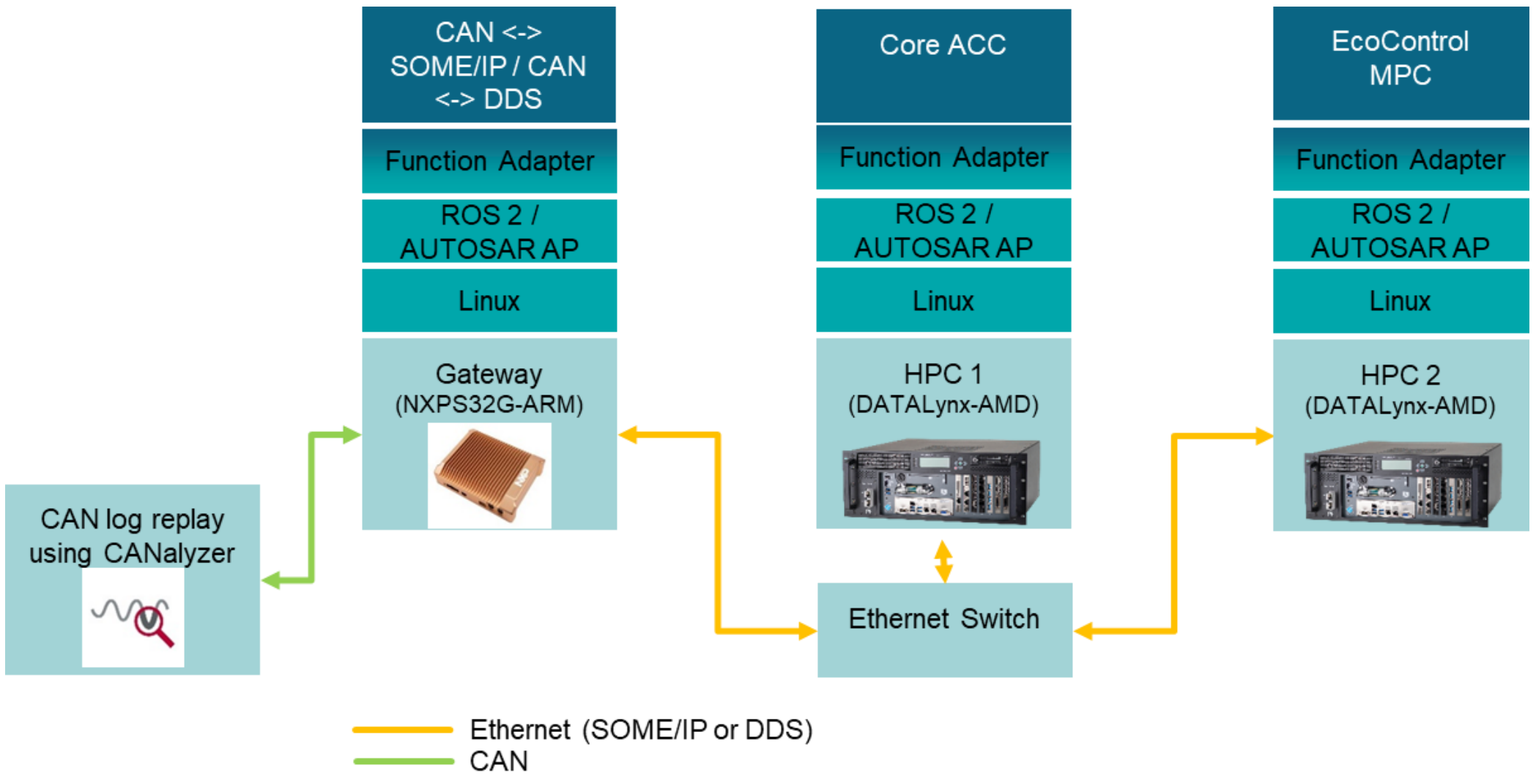

*Figure 10: Overall architecture of the EcoControl application integrated onto different software platforms (ROS 2 and Adaptive AUTOSAR).*

## 5.2 Application overview

An overview of the EcoControl application with its corresponding software components and high-level interfaces is presented in the Figure 11. It contains two software components namely Core Adaptive Cruise Control (Core ACC) and Model Predictive Control (MPC) with a cycle time of 50ms and 500ms, respectively. The Core ACC determines the acceleration request for the powertrain based on a multitude of factors such as the set speed by the driver, distance to the preceding vehicle in the same lane and the curvature of the lane. The MPC is an energy saving functionality that computes the acceleration request based on the altitude of the horizon. The output of the MPC is then arbitrated in the Core ACC component to compute the final acceleration request. The EcoControl application is modelled in Simulink and the C code is generated from the model using the Simulink Coder. Additionally, the data interfaces of the EcoControl application are standardized according to the Vehicle Signal Specification (VSS). An exemplary visualization of the VSS tree for the EcoControl application with some selected signals can be seen in the Figure 12.

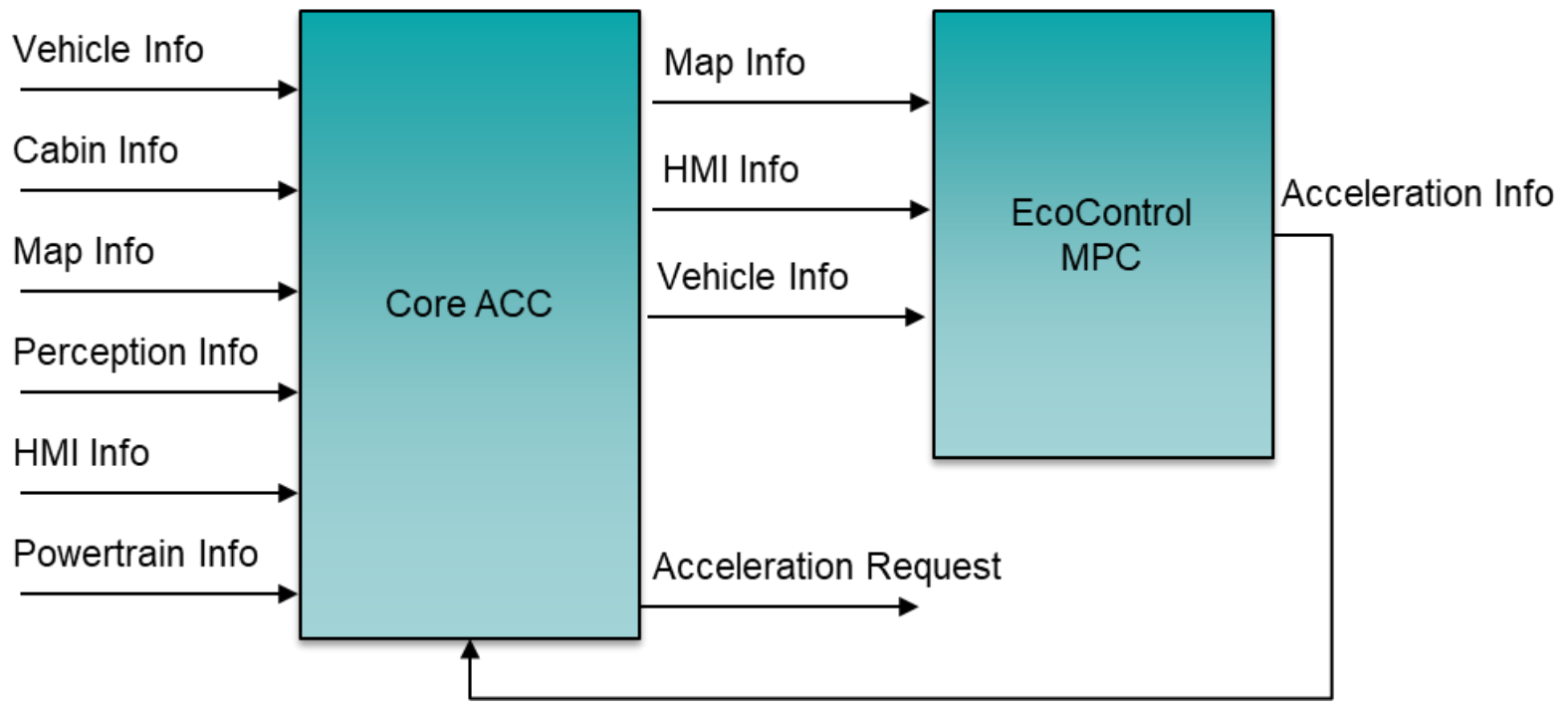

*Figure 11: Components and high-level data interfaces of EcoControl application: Flowchart showing data exchange between Core ACC and EcoControl MPC systems.*

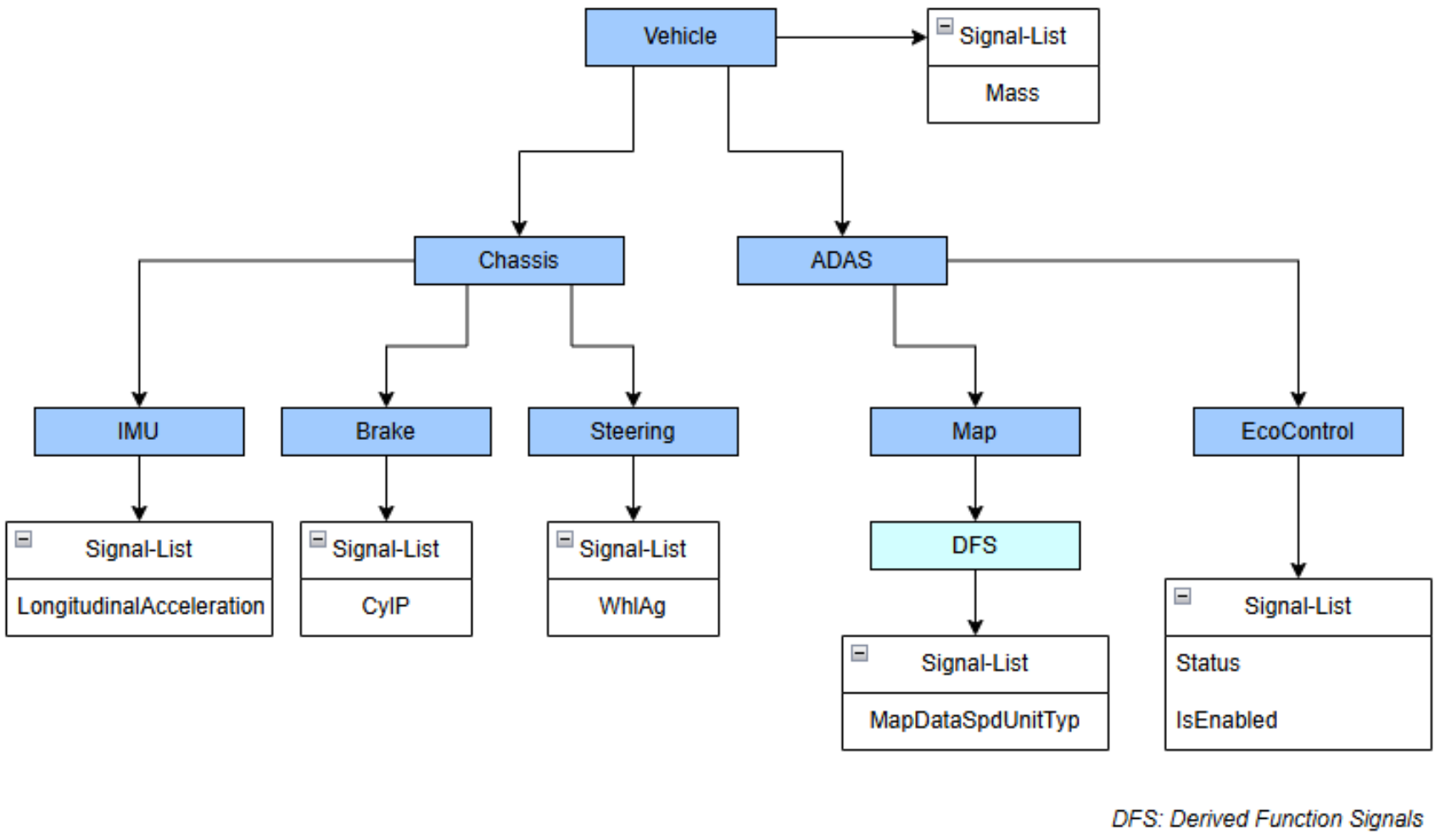

*Figure 12: VSS tree view for the EcoControl application with some example signals.*

The *function model* and the *integration model* for both Adaptive AUTOSAR and ROS 2 are generated to enable efficient integration of the EcoControl application onto the software platforms.

## 5.3 Under development

This section outlines both the implemented and ongoing aspects of the workflow. Currently, artefacts such as the *function* and *integration JSON* models are available. However, the mapping between them is performed manually. A key future objective is to enable the automatic generation of the *integration JSON* from the *function JSON* for a given software framework.

The concept of a *Platform Function*, which involves separating an application into project-specific and core-specific modules, is outside the scope of this use case and has therefore not been implemented. Additionally, the generation of *function adapter* code for the ROS 2 implementation is currently a manual process. Automating this step is a future goal aimed at improving integration efficiency. In contrast, this step is already automated in the case of Adaptive AP.

### 5.4 Demonstrate concept on Adaptive AUTOSAR

In the case of Adaptive AUTOSAR, applications utilize the AUTOSAR Runtime for Adaptive Applications (ARA). This runtime environment provides interfaces for the integration of different applications into the system. ARA offers mechanisms for ECU-internal and inter-network communications as well as access to basic services such as diagnostics, network management, persistent saving of data, functional monitoring of the platform, logging of measurement values, update and configuration management, etc. An adaptive application is typically implemented as at least one process in a POSIX-based operating system. Each adaptive application consists of an executable program and a manifest file, together forming a package.

For any Adaptive AUTOSAR project, it is necessary to configure and manage the software components so that it can be deployed onto an HPC. In this work, DaVinci Developer Adaptive, a tool from Vector Informatik GmbH, is used to configure the software components. Typically, this process is carried out manually and requires considerable effort. However, with all the application and software platform related information available in the *integration model*, this information is then used alongside a *model configuration tool* to automatically configure the project in DaVinci Developer Adaptive. This step produces ARXML files related to the application, data types, service interfaces and deployment catalogues.

The EcoControl application code, generated using the Simulink coder, contains entry-point functions such as `initialize`, `step`, and `terminate`, which aligns with our established *Platform Function* concept. These functions exchange input and output through global variables. This data and the exchange mechanism are the interfaces of the entry-point functions. However, in the case of Adaptive AUTOSAR applications, the data transmission is service-oriented. This means that the SOME/IP messages received over the ethernet network are deserialized by the SOME/IP daemon of Adaptive AUTOSAR and the deserialized data is then used to invoke the corresponding service. The `ara::com` APIs can then be used to access the variables of the services. Currently, the `ara::com` APIs for the subscription and publication of services should be manually coded and a manual mapping of the `ara::com` APIs to the global variables of the EcoControl application is required. This step is time consuming, particularly when the number of interfaces to the application is high. This work increases efficiency by using the Application Framework tool [15] from Vector Informatik GmbH to automatically generate skeleton code for the adaptive application and the `ara::com` APIs. The input to the Application Framework is also the JSON file generated from the integration model. It is also important to note that the APIs generated by the Application Framework tool are standardized based on the VSS. An overview of the workflow is shown in the Figure 13. This ensures that instead of manually writing the APIs, it is generated automatically, thereby enabling faster integration. The `SetInput` function is then responsible for reading the data through `ara::com` subscription APIs and map it to the input global variables of the application. Similarly, `GetOutput` function maps the output global variables to the `ara::com` publication APIs. The `Step` function is where the main application logic is executed. It is important to note that the APIs generated by the Application Framework is analogous to the concept of Function Adapter described earlier.

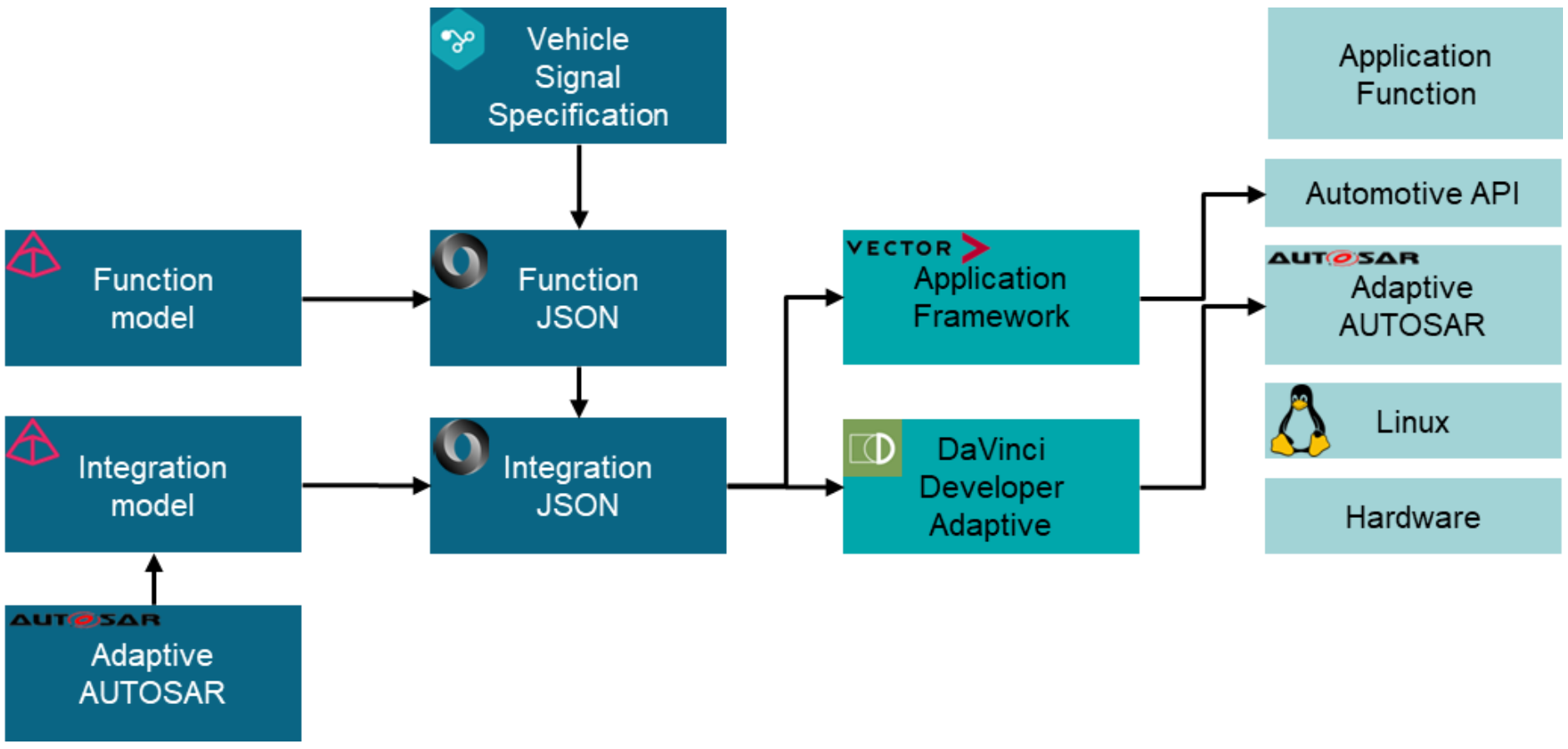

*Figure 13: Workflow for integration of applications onto Adaptive AUTOSAR: Flowchart showing the process of vehicle signal specification integration.*

## 5.5 Demonstrate concept on ROS 2

In ROS 2, the fundamental building blocks of a system are referred to as nodes. A node can instantiate any number of subscribers and publishers for communication. To achieve maximum flexibility in the execution of nodes, it was decided that the ROS 2 function adapter would be a ROS 2 lifecycle node. This means that in the implementation, the ROS 2 function adapter inherits from the ROS 2 lifecycle node. The advantage of using a ROS 2 lifecycle node is that the execution state can be precisely specified and controlled. This would allow for further development in terms of operational mode management. The state machine which defines the states and transitions of the ROS 2 lifecycle nodes can be seen in Figure 14**Fehler! Verweisquelle konnte nicht gefunden werden.**.

During the transitions "Configuring" and "ShuttingDown," the APIs "Init" and "Terminate" of the platform function are called, respectively. The functions "setExternalInputs," "Step," and "getExternalOutputs" are executed if the node is in the "Active" state.

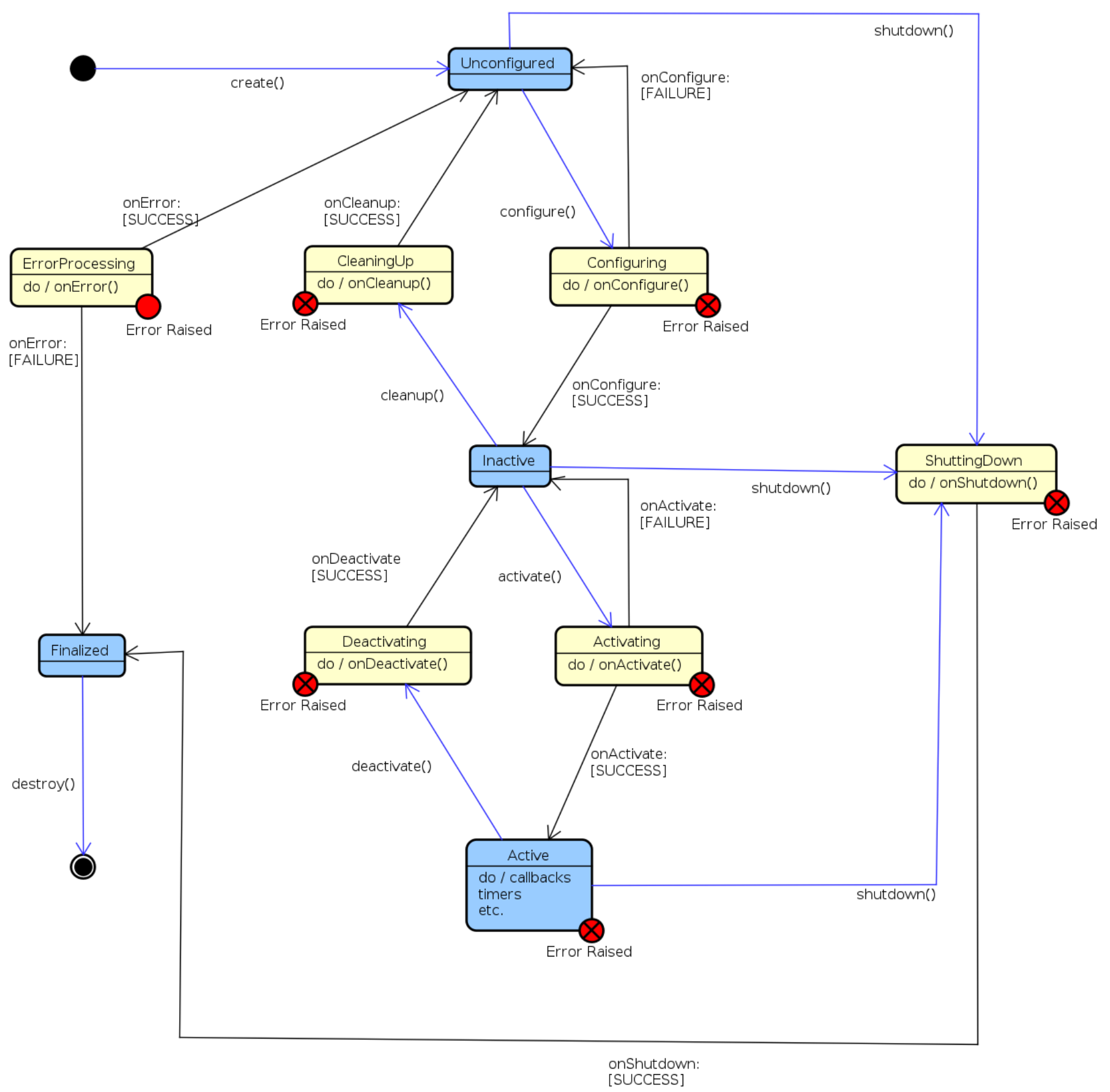

*Figure 14: Lifecycle of a ROS 2 node as a series of states and transitions. [16].*

Receiving new data via a subscriber always occurs event-based and cannot be blocked without further configurations. This necessitates storing incoming messages in local buffer variables. The publisher methods can be configured to be called cyclically. In a publisher method, the three execution steps of the platform function are then carried out. The methods "setExternalInputs" and "getExternalOutputs" must be used to convert the data from the local function adapter variables into the global variables of the platform function. The sequence diagram in Figure 15 illustrates how the function adapter moves through the ROS 2 lifecycle states and makes clear when the APIs of the Platform Function are called.

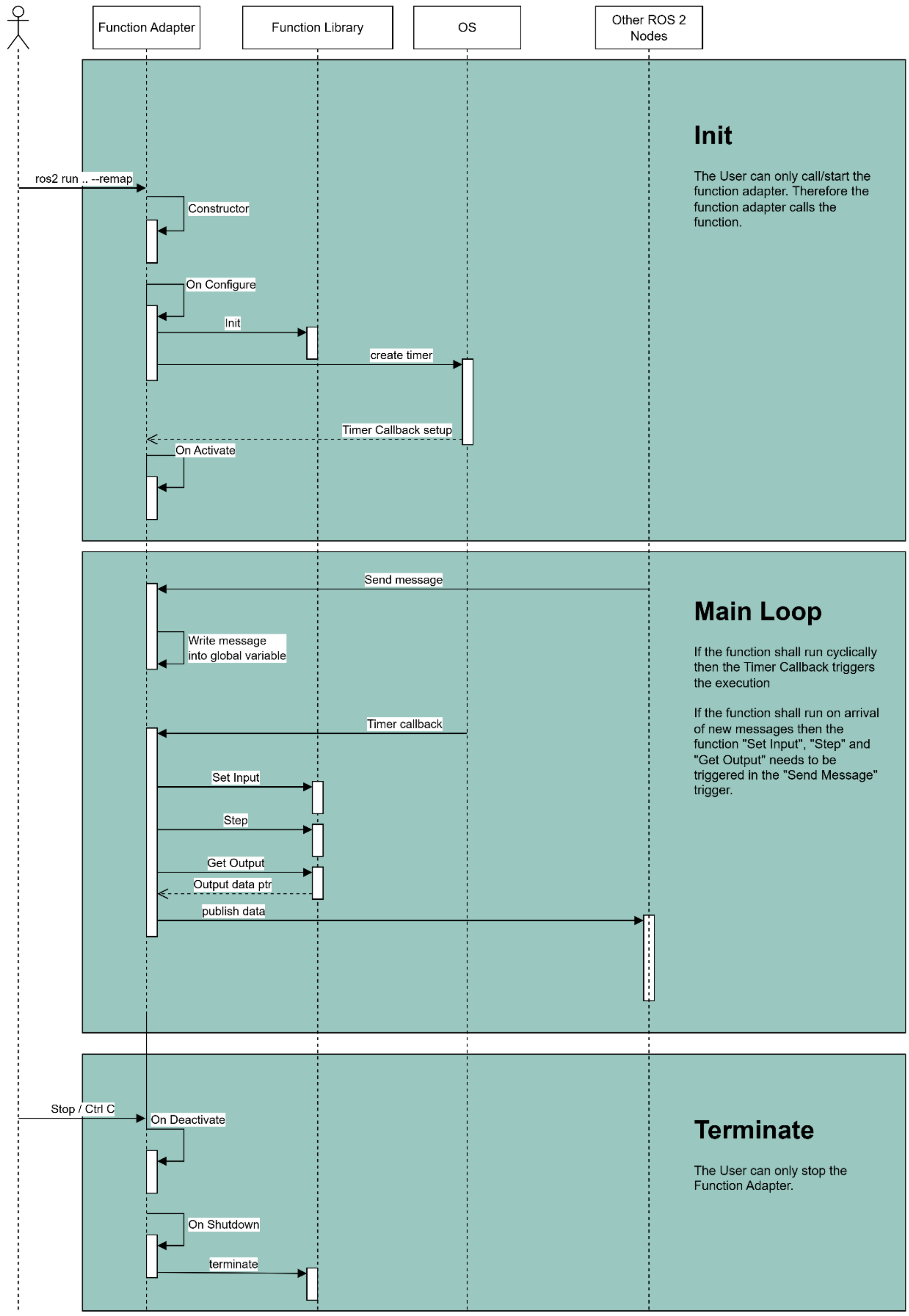

*Figure 15: Sequence diagram of the function adapters lifecycle. It starts in the Init phase and then moves to the Main loop where the business logic of the platform function is executed. Once the Function Adapter is stopped the Terminate Sequence is executed.*

## 5.6 KPI Definitions

This section presents the results of the integration and the derived Key Performance Indicators (KPIs) that demonstrate the performance and efficiency of the approach. The KPIs are compared across three implementations: one with the ROS 2 implementation without the *Platform Function* concept (baseline), two with the ROS 2 implementation with the Platform Function concept and three with the *Platform Function* concept applied to Adaptive AP.

The KPIs are categorized into three:

1. *Function behavior KPIs* – Metrics to compare the behavior of the application integrated on Adaptive AUTOSAR and ROS 2 using our proposed method against the baseline.
2. *Timing and load KPIs* – Metrics to evaluate the timing and load aspects of the application and the function adapter.
3. *Integration efficiency KPIs* – Metrics to show the integration efficiency of the approach.

The open loop setup described earlier is used for the derivation of the metrics.

### 5.6.1 Function behavior

To ensure successful integration, it is essential to verify that the application's behavior remains consistent with the baseline. Traces from the Baseline, ROS 2, and Adaptive AUTOSAR (Adaptive AP) integration approaches were captured using synchronized logging mechanisms and analyzed for both temporal alignment and value consistency.

For this analysis, two key services were selected: the ego vehicle longitudinal acceleration (an input to the application) and the Adaptive Cruise Control (ACC) internal status (an output of the application). These services were examined across all three implementations, as illustrated in Figure 16 (input) and Figure 17 (output). In these figures, the x-axis represents the timestamps, while the y-axis shows the corresponding service values.

Across all test runs, no measurable differences were observed between the three approaches. By comparing the input at the Core ACC component, we confirm that the CAN messages replayed from the CANalyzer log are correctly decoded by the Gateway, published via the appropriate middleware SOME/IP for Adaptive AP and DDS for ROS 2 and successfully subscribed to by DATALynx1.

Furthermore, the comparison of the output service validates that both the Core ACC and MPC components are functioning as expected, and that the communication between DATALynx1 and DATALynx2 through the middleware is operating correctly. These results demonstrate that the abstraction introduced by our approach does not compromise runtime correctness or control fidelity. The observed equivalence in service behavior confirms the robustness of the standardized interface concept and supports its applicability for cross-platform deployment.

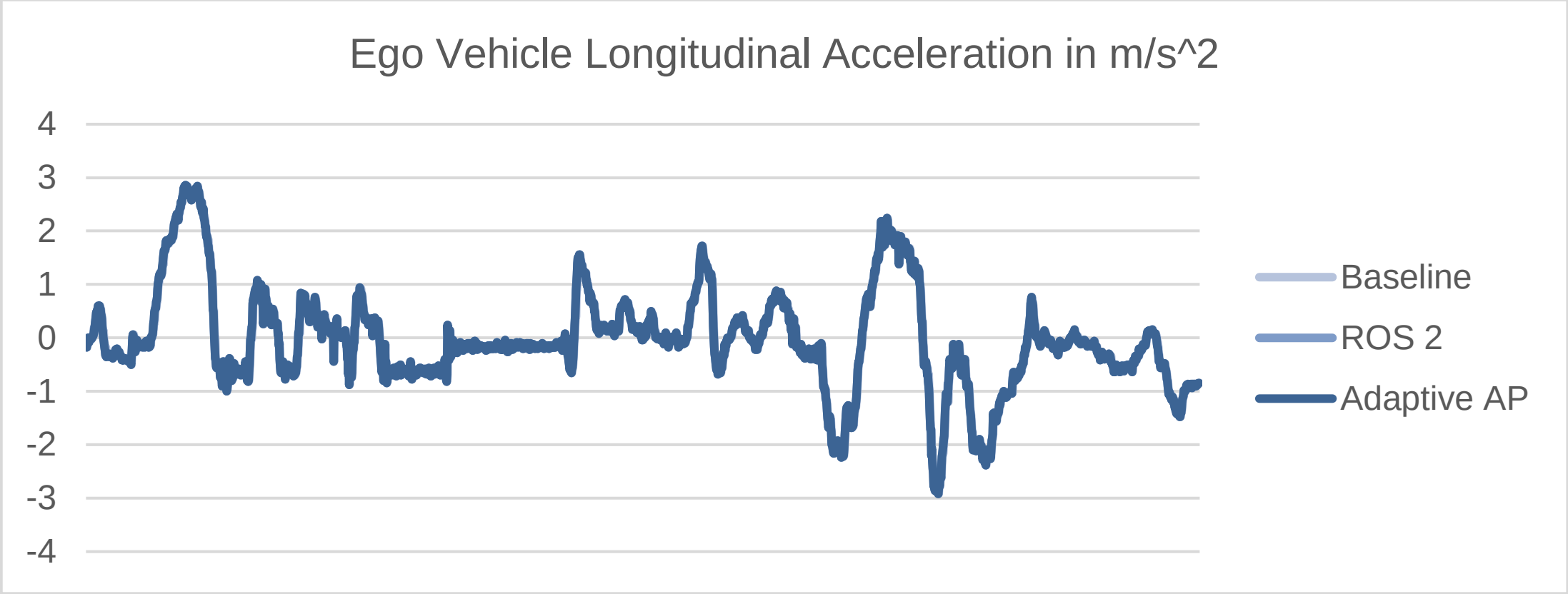

*Figure 16: Comparison of the input to the Core ACC component across three implementations.*

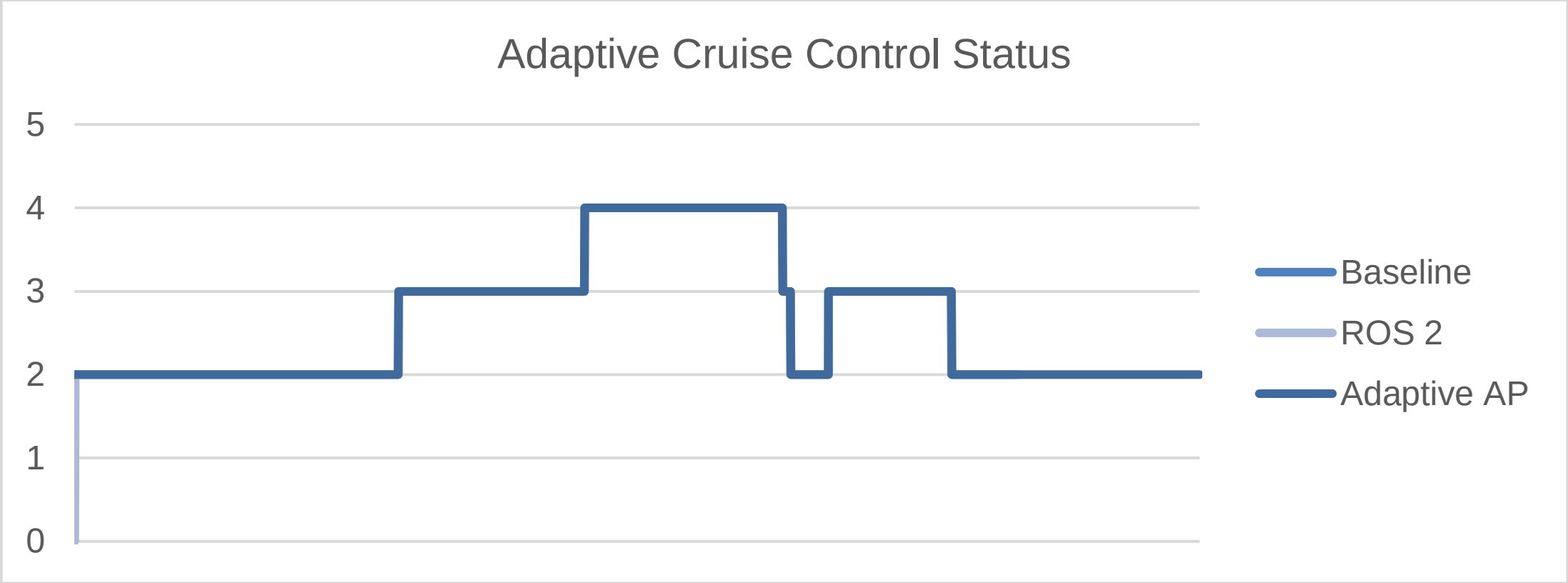

*Figure 17: Comparison of the output of the Core ACC component across three implementations.*

### 5.6.2 Timing and load

The aim of these metrics is to ensure that the timing requirements of the components are consistently met, and that the introduction of our approach does not impose additional computational overhead—specifically in terms of CPU load or RAM consumption. Hence, the following metrics are introduced namely with respect to the timing aspects:

- Total execution time of the components ($t_{\text{exec}}$)
- Time taken by the function adapter layer ($t_{\text{adapter}}$)
- Time taken for the function logic ($t_{\text{logic}}$)

With regards to the computational aspects, following metrics are introduced:

- CPU usage
- RAM usage
- Executables size

Top left section of the Figure 18 presents the execution times associated with the CAN-to-middleware translation across three platforms: Adaptive AUTOSAR (CAN to SOME/IP), ROS 2, and the baseline (both using CAN to DDS). It is the time required to read CAN messages from Socket CAN and subsequently publish a single service using the respective middleware on the Goldbox hardware. The median execution times observed are 434 µs for the baseline, 423 µs for ROS 2, and 302 µs for Adaptive AUTOSAR. Although maximum execution times occasionally reach up to 23,000 µs, these remain well within the cyclic execution window of 50 ms, thereby ensuring reliable system performance.

The top-right section of Figure 18 shows the execution time of the Core ACC logic across all three platforms. With median values around 100 µs, it is evident that the platforms themselves do not introduce any significant computational overhead. In contrast, the execution time of the MPC logic, shown in the middle-left section of Figure 18, reveals a notably higher median value for Adaptive AUTOSAR compared to ROS 2 and the baseline. However, given the MPC's cycle time of 500 ms, this increased execution time remains acceptable, and further optimization was considered beyond the scope of this study.

To evaluate the impact of the platform function concept, the execution time of the function adapter was also analyzed for both Core ACC and MPC components across ROS 2 and Adaptive AUTOSAR. The middle-right section of Figure 18 presents the function adapter execution time for Core ACC, with a median value of approximately 300 µs across both platforms, indicating minimal overhead. A similar trend is observed for the MPC function adapter with a median value of around 200µs, as shown in the bottom section of the Figure 18.

These results demonstrate that the evaluated platforms are capable of meeting the timing requirements without introducing significant computational overhead. The consistency in execution times across components and platforms supports the feasibility of the proposed platform function concept for reliable deployment.

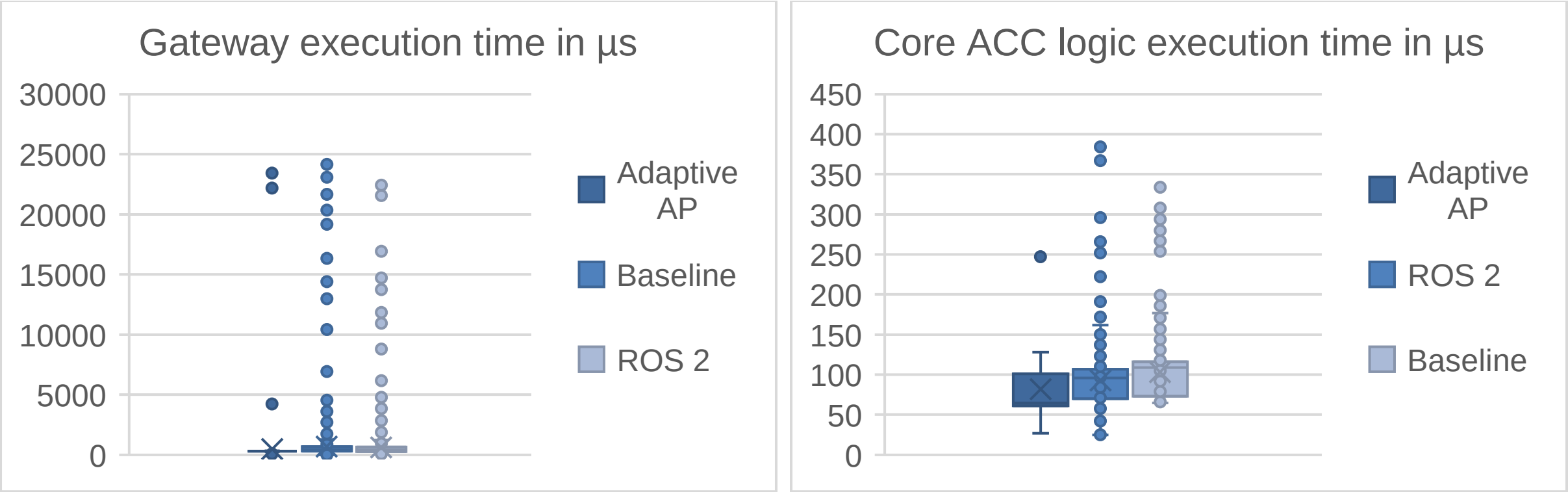

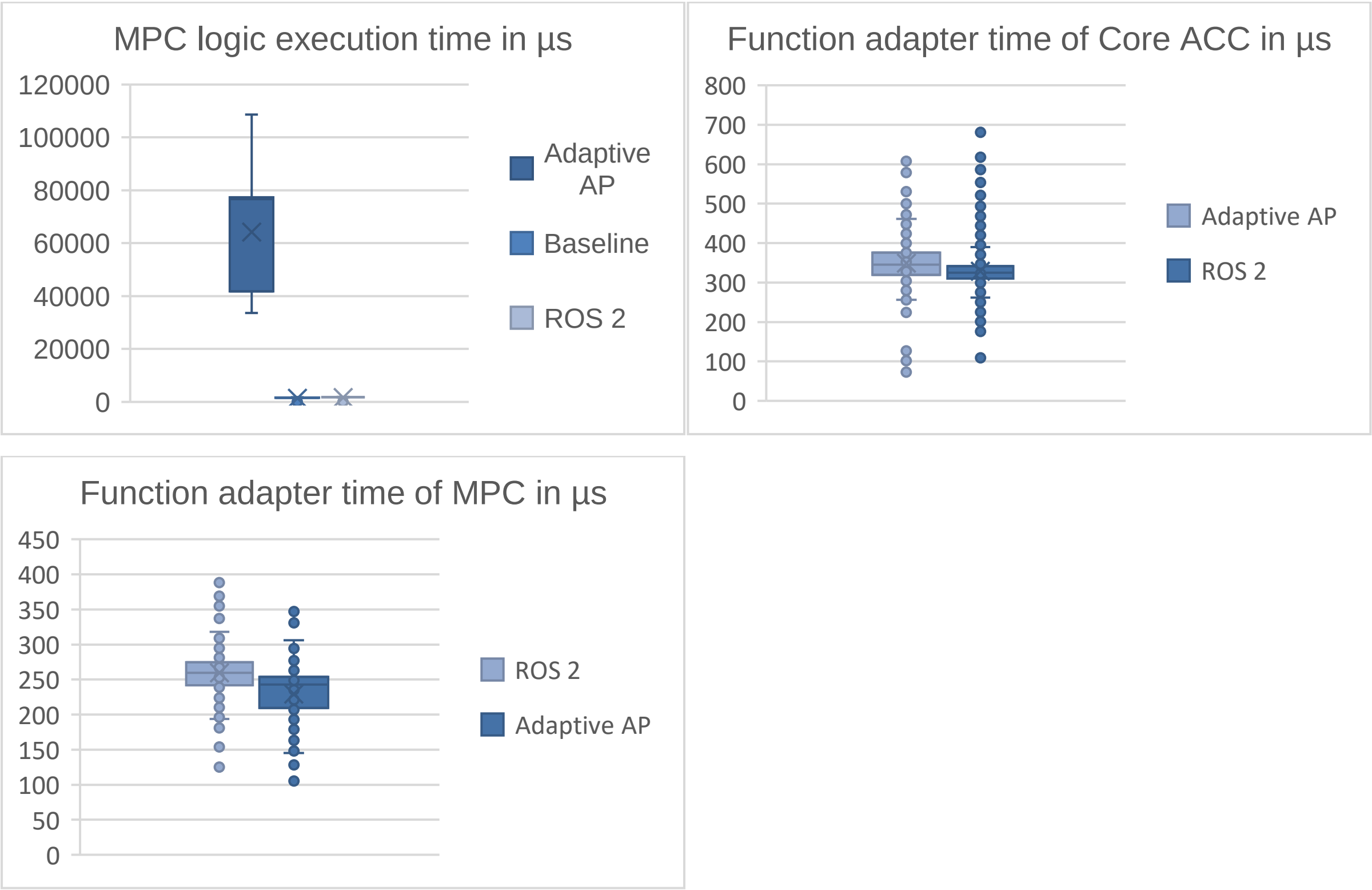

*Figure 18: Timing metrics involving the gateway, application logic and function adapter across three platforms.*

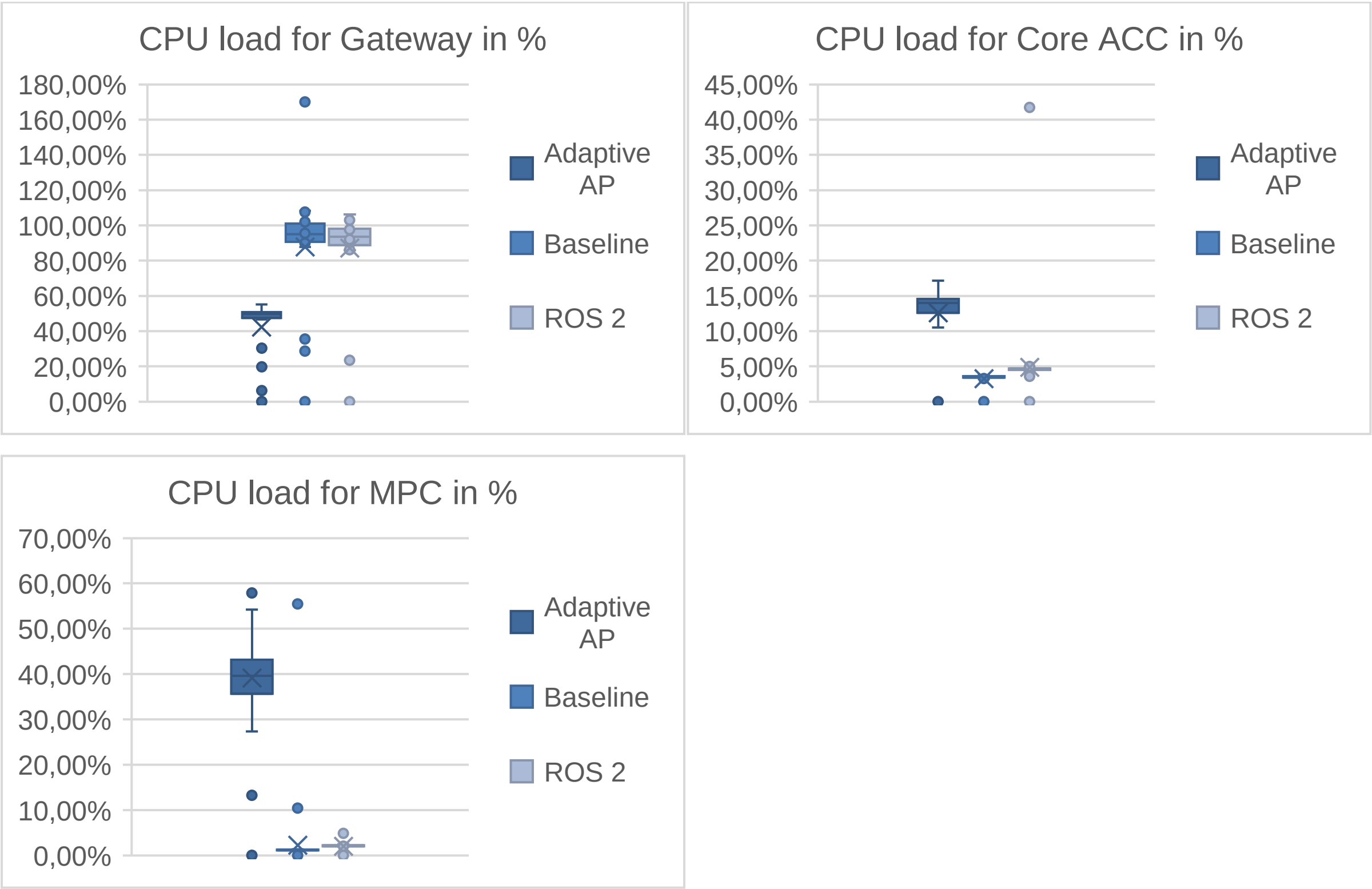

*Figure 19: CPU load metrics for Gateway, Core ACC and MPC across three platforms.*

Figure 19 presents a comparative analysis of CPU load across different hardware platforms and software implementations. The top left part of the Figure 19 shows the CPU utilization of the CAN to SOME/IP executable for the Adaptive AUTOSAR Platform (Adaptive AP), and the CAN to DDS executable for both the baseline and ROS 2 implementations on Goldbox hardware. Notably, the median CPU load for the baseline and ROS 2 implementations approaches 100%, whereas the Adaptive AP implementation exhibits a significantly lower median load of approximately 50%. Given that the Goldbox hardware comprises four cores, this indicates not a significant resource consumption.

In the top-right section of Figure 19, the CPU load of the Core ACC executable on DATALynx1 hardware, which features 20 cores, is depicted. Here, the baseline and ROS 2 implementations maintain a median CPU load of around 5%, while the Adaptive AP implementation shows a higher median load of approximately 15%.

The bottom section of the Figure 19 highlights the CPU load for the MPC executable on DATALynx2 hardware. The Adaptive AP implementation demonstrates a notably higher median CPU load of about 40%, in contrast to the baseline and ROS 2 implementations, which remain at around 4%.

Although the elevated CPU usage observed in the Adaptive AP implementations is beyond the scope of this study, it is important to note that the *Platform Function* concept itself does not contribute to this computational overhead. Instead, the increased load is attributed to the underlying platform characteristics of Adaptive AUTOSAR.

Figure 20 illustrates the memory usage of various executables, each deployed within dedicated containers and allocated 32 GB of RAM. The top-left section of the Figure 20 presents the RAM consumption of the Gateway executable, which averages between 0.5% and 2% across the three evaluated platforms. The top-right section of the Figure 20 shows the RAM usage of the Core ACC executable, consistently around 0.2% regardless of the platform. The bottom section of the Figure 20 depicts the RAM consumption of the MPC executable, which remains approximately 0.15% across all platforms.

These results indicate that memory usage remains minimal and consistent across implementations. Furthermore, the data supports the conclusion that the *Platform Function* concept does not introduce additional overhead in terms of RAM consumption.

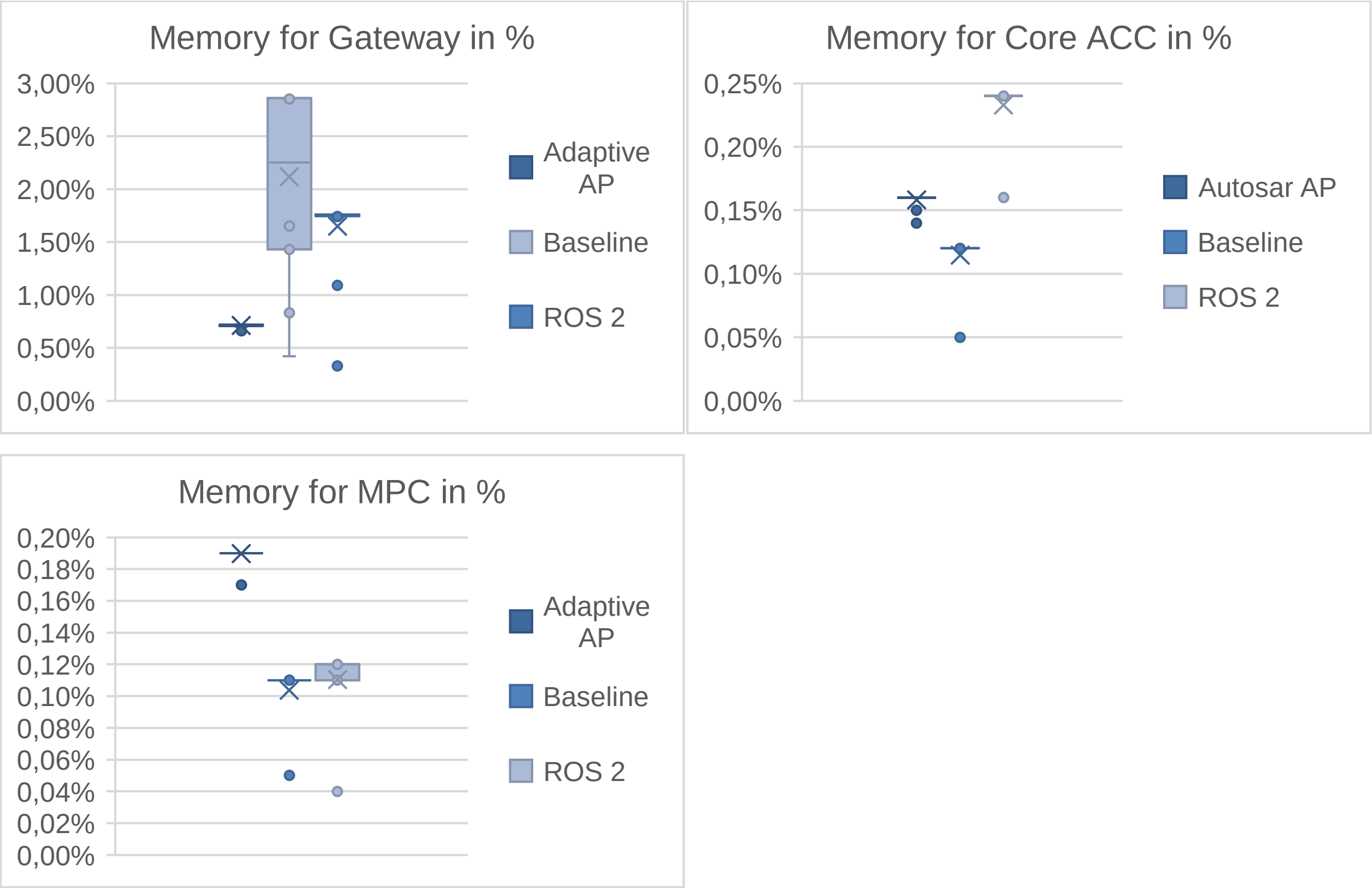

*Figure 20: Memory consumption in % for Gateway, Core ACC, and MPC across three platforms.*

### 5.6.3 Integration Efficiency

To quantify the efficiency gains achieved through our *Platform Function* approach, we analyzed the implementation effort required for the Function Adapter across the ROS 2 and Adaptive AP platforms. In the ROS 2 implementation, the Function Adapter codebase consists of 600 Lines of Code (LoC) for MPC and Core ACC 1250 LoC. This increase reflects the additional abstraction and modularization introduced to support standardized interfaces and reusable integration logic. However, despite the larger codebase, the Platform Function approach significantly improves integration efficiency due to its high potential for automation.

Based on a detailed analysis of the code structure, we estimate that approximately 95% of the Function Adapter code in the Platform Function variant can be automatically generated using model-driven tools and standardized interface definitions. This translates to 570 LoC of the 600 LoC for MPC 1187 LoC of the 1250 LoC for Core ACC. Assuming a conservative estimate of 1 minute per LoC for manual implementation (excluding design and debugging effort), the potential time savings through auto generation are substantial:

- MPC: ~570 minutes (≈9.5 hours) saved
- Core ACC: ~1187 minutes (≈19.8 hours) saved

In the Adaptive AUTOSAR development process, project configuration involves defining executables, services, events, and other system elements. This is typically performed using the DaVinci Developer Adaptive tool provided by Vector Informatik GmbH. However, the configuration process is manual and time-consuming. For instance, in our project, which comprises 3 executables, 18 services, and 162 events, manual configuration requires approximately three hours.

To address this inefficiency, we utilize the *Model Configurator tool*, which automates the configuration process by consuming the *Integration JSON* file, generated for the EcoControl application as described in the beginning of this chapter, as input. This approach enables the automatic generation of the Adaptive AUTOSAR project within DaVinci Developer Adaptive, reducing the configuration time to less than one minute. This significantly improves integration efficiency and minimizes the potential for human error during setup.

The Application Framework significantly enhances integration efficiency by automating the generation of code required to bind application logic with the `ara::com` APIs. In our project, which includes 162 events across the Core ACC and MPC executables, this automation results in approximately 3,500 lines of code (LoC). Manual implementation of this code would typically require several days of integration effort. In contrast, the automated approach completes this task in under 30 minutes, representing a substantial reduction in integration time.

Currently, the mapping of global application variables to the application framework APIs remains a manual process, accounting for an additional 1,000 LoC. Automating this step presents a clear opportunity for further optimization.

These figures underscore the practical advantages of adopting a standardized, tool-supported integration strategy. By minimizing manual coding effort, the *Platform Function* approach not only accelerates development but also reduces the likelihood of human error and enhances maintainability across platforms.

# 6 Conclusion

With the work we have shown the applicability of Function Model and Integration Model in combination with industrial grade tools. The introduced function adaptor code between platform function and base software didn’t show a significant impact on the runtime performance and resource utilization. Following the VSS signal approach alongside API standardization makes the road ready for improved efficiency in function integration. Nonetheless the introduction of containerized workloads opens flexibility in the deployment of functions. With the concept we have shown that the effort in integration of functions can be reduced via automation and there is enough headroom to gain further benefits from automation.

## 7 Acknowledgement

This work was partly supported by the German Federal Ministry of Education and Research (BMBF) as part of the project autotech.agil (FKZ 01IS22088).

Furthermore, we would like to thank the outstanding support of our colleagues at ZF Technology Centre India to improve the overall concept and transport it into further applications.

This paper was originally presented at the ELIV Conference 2025 in Bonn, Germany, and published in the corresponding VDI Report.

## 9 Glossary

**Error Interface**: These interfaces manage error-related interactions, allowing the application to send error log requests or error reset requests to the failure or fault management components within the BSW. This standardization ensures consistent error handling and logging practices across platforms, facilitating clear communication of fault data. Naming convention for these interfaces shall be as shown below.

`FunctionName_ErrorName_ErrorSts`

E.g.: WheelSpeedCalculation_RangeError_ErrorSts

**Safety Reaction Interface**: These interfaces communicate critical information from the BSW's failure management system to the application layer, alerting it to errors or safety-related events. This interface category provides a structured channel for safety signals, ensuring that applications can respond appropriately to safety situations according to standardized protocols. Naming convention for these interfaces shall be as shown below.

`<SafetyCondition>_SftyCondSts`

E.g.: WhlSpdRLInvalid_SftyCondSts

**Mode Management Interface**: These interfaces convey system or functional mode information essential for coordinating application behaviour across different operational states. System modes or functional modes are typically influenced by various system conditions, and these interfaces enable applications to adapt accordingly, maintaining functionality across diverse modes through a unified format. Naming convention for these interfaces shall be as shown below.

`<FunctionName>_Mode_In <FunctionName>_Mode_Out`

E.g.: WheelSpeedCalculation_Mode_In(), WheelSpeedCalculation_Mode_Out()

**Scheduling Interfaces**: Scheduling interfaces differ significantly from the other categories, as they are not signal based but instead implemented as function calls. This design enables scheduling interfaces to manage procedural operations like initialization, main runtime, and shutdown processes within the application, which require precise timing and sequencing. Additionally, when safety-critical requirements demand, these function calls can also indicate the need for watchdog supervision in BSW to monitor timing and safety, ensuring that the system meets strict operational and safety standards. Further, they can also carry memory requirements of the function that may be considered during integration wherever needed. Naming convention for these interfaces shall be as shown below.

```
void <FunctionName>_init()
void <FunctionName>_Step()
void <FunctionName>_Terminate()
```

E.g.: WheelSpeedCalculation_Init(),

WheelSpeedCalculation_Step(),

WheelSpeedCalculation_Terminate()

| Interface type | Fields | Description of the fields |
|---|---|---|
| Data/ Parameter (As per VSS with extension to fields) | Type | Indicates if it is a data interface or a parameter interface |
| | Default | Default value of the interface |
| | Datatype | Datatype (float, uint8, uint16 etc.) |
| | Min | Minimum limit of interface range |
| | Max | Maximum limit of interface range |
| | Unit | Unit detail |
| | Description | Brief overview on the interface: How is it calculated, relevant to which sensor/ actuator/ domain/ method etc. |
| | ASIL | ASIL level (QM/A/B/C/D) |
| Error Info | Datatype | Datatype of the Error Signal |
| | Maturation time | Duration for which the Error condition must be valid, for the error to be logged (ms) |
| | Severity | Description of severity: Disable entire function, partial degradation, set warning lamp etc. |
| | Reset time | Duration after which the error shall be reset after detecting an error reset condition (ms) |
| | Reset condition | Description of conditions that shall reset and existing error. |
| | Description | Description of the error: conditions, scenarios, etc. |
| | Dependency (Optional) | Error's' from other functions or BSW due to which this error is also set automatically |
| Safety Reaction | Datatype | Data type of the Signal |
| | Error list | List of Errors which can cause this safety condition, if any of the error in the list is set. |
| | Description | Description of the safety Reaction |
| Modes | Datatype | Data type of the interface |
| Scheduling init() | Description | Description of init function |
| Scheduling step()<br><br>includes Watchdog & Allocation | Run type | Run type of the function like init/cyclic etc. |
| | Cycle time | Periodicity of the function execution in milliseconds |
| | Description | Description of the Step function |
| | Initial execution offset (Optional) | Offset in milliseconds of the Step function before its execution |
| | Priority (Optional) | Relative priority of the Step function w.r.t other Step functions of the Platform Function/Feature |
| | Function Scheduling (Optional) | Scheduling mechanism as description: Pre-emptive/Non-Pre-emptive etc. |
| | Debounce time (Optional) | Mandatory time gap in milliseconds for the next execution of the same Step function. |
| | Implemented ASIL Level | ASIL Level for which the Step function to be implemented |
| | Previous runnable (Optional) | Previous Step function after which this Step function must be executed. |
| | Supervision type | Type of Software Watchdog Supervision |
| | Alive limits (Optional) | Alive Supervision Limits like Min Indications, Max Indications, Reference Indications and Error Name. Description if needed |
| | Deadline limits (Optional) | Deadline Supervision Limits like Min duration and Max duration in milliseconds allowed for the execution |
| | | time of the function, Error Name. Description if needed |
| | Logical check (Optional) | Logical Supervision hint like, description of the order w.r.t other functions of the feature, Error Name. Description if needed |
| | Stack size (Optional) | Stack size in bytes |
| Scheduling terminate() | Description | Description of terminate function |